\documentclass[11pt]{article}
\usepackage{amssymb,graphicx,epsfig,cite,color}
\renewcommand\baselinestretch{1.15}

\begin{document}
\thispagestyle{empty}

OSU-HEP-13-08 \hfill TIFR-TH/13-27

\bigskip

\begin{center}
{\LARGE\bf Higgs Boson Decay Constraints on a Model \\ [2mm]
with a Universal Extra Dimension}

\bigskip

{\large\sl Anindya Datta}\,$^{a,1}$,
{\large\sl Ayon Patra}\,$^{b,2}$
and
{\large\sl Sreerup Raychaudhuri}\,$^{c,3}$
 
\bigskip 
 
{\small
$^a$ Department of Physics, University of Calcutta, \\
92 Acharya Prafulla Chandra Road, Kolkata 700\,009, India \\ [2.5mm]
$^b$ Department of Physics, Oklahoma State University, 
Stillwater, Oklahoma OK 74078, U.S.A. \\ [2.5mm]
$^c$ Department of Theoretical Physics, Tata Institute of Fundamental 
Research, \\ 1 Homi Bhabha Road, Mumbai 400005, India.
}
\end{center}

\bigskip\bigskip

\begin{center} {\large\bf Abstract} \end{center}
\vspace*{-0.35in}
\begin{quotation}
\noindent We investigate the impact of the latest data on Higgs boson 
branching ratios on the minimal model with a Universal Extra Dimension 
(mUED). Combining constraints from vacuum stability requirements with 
these branching ratio measurements we are able to make realistic 
predictions for the signal strengths in this model. We use these to find 
a lower bound of 1.3~TeV on the size parameter $R^{-1}$ of the model at 
95\% confidence level, which is far more stringent than any other 
reliable bound obtained till now.
\end{quotation}

\bigskip

\centerline{\sf Pacs Nos: 14.80.Bn, 14.80.Rt, 11.10.Kk}

\vfill

\centerline{\today}

\bigskip

\hrule
\vspace*{-0.1in}
$^1$ adphys@caluniv.ac.in \hspace*{0.8in} 
$^2$ayon@okstate.edu \hfill 
$^3$ sreerup@theory.tifr.res.in

\newpage

The discovery of the 125-126~GeV Higgs boson --- or its close lookalike 
--- at CERN, Geneva, in the previous year~\cite{Higgs_discovery}, has 
proved to be a game-changing moment in phenomenological studies of 
electroweak interactions. Gone are speculations about Higgsless 
models~\cite{Higgsless}, strongly-coupled Higgs 
sectors~\cite{strongly_coupled} and fears that the Higgs boson 
self-coupling may hit a Landau pole at some large energy 
scale~\cite{Landau_pole}. Instead, today's theoretical studies have 
other concerns, such as stability of the electroweak vacuum, fine-tuning 
constraints and the requirement that the measured Higgs boson mass and 
branching ratios be correctly explained in whatever model happens to be 
the subject of the study. At the present instance, there is no 
compelling reason, beyond certain theoretical prejudices (like grand 
unification), to believe that we require anything other than the 
Standard Model (SM) to explain all the known phenomena on a terrestrial 
scale. Destabilisation of the SM vacuum at some energy scale below the 
Planck scale could be one of the strongest hints of new 
physics~\cite{vacuum_stability}, but at the moment this issue is mired 
in uncertainties of the top quark mass measurement~\cite{top_mess}.

Nevertheless, we do require physics beyond the Standard Model, and this 
requirement arises as soon as we look outside the confines of our Earth 
into the cosmos beyond. Here it is well known that the SM fails to 
provide explanations for ($i$) the composition of dark 
matter~\cite{dark_matter}, ($ii$) the nature of dark 
energy~\cite{dark_energy} and ($iii$) the amount of $CP$-violation 
required for baryogenesis~\cite{CP_violation}. Of these, perhaps the 
most tractable problem is the first one, viz. the generation of a model 
for dark matter, for all that is required is a model for a stable, 
weakly-interacting massive particle (WIMP). The most famous model which 
provides this is, of course, supersymmetry with conservation of 
$R$-parity, where the lightest supersymmetric particle is the WIMP in 
question~\cite{SUSY}. An alternative model, which was proposed about a 
decade ago, is one with a so-called Universal Extra 
Dimension~\cite{Appelquist}. In the minimal model of this kind (mUED), 
each five-dimensional SM field is replaced by a tower of Kaluza-Klein 
(KK) modes, each labelled by a KK number $n$, and having masses given 
(at tree-level) by $M_n = \left(M_0^2 + n^2 R^{-2}\right)^{1/2}$. Here, 
the lightest of the $n = 1$ particles is stable and weakly-interacting 
due to a $Z_2$ symmetry called KK parity, defined in terms of KK number 
by $(-1)^n$. This lightest KK particle, called the LKP, is an excellent 
candidate for dark matter~\cite{Hooper_Profumo}.

At a high energy collider, the behaviour of the mUED models is very 
similar to that of supersymmetric models~\cite{Matchev}. The $n = 1$ 
states form analogues of the supersymmetric particles, exhibiting 
cascade decays ending in the LKP, which is then a source of missing 
energy and momentum. A major difference from supersymmetry is the 
presence of $n =2$ and higher KK modes, which could perhaps be produced 
as resonances in a high energy machine like the LHC~\cite{resonances}. 
However, a more significant difference arises when we consider the 
ultraviolet behaviour of the mUED model (or any model with KK modes), as 
was pointed out in a pioneering paper by Dienes {\it et 
al}~\cite{Gherghetta}. This is the fact that when we allow the SM 
coupling constants to run in this model, we encounter repeated KK 
thresholds at every scale $n/R$, so that, when considered over a large 
range of energies, the coupling constant exhibits a piecewise 
logarithmic running closely mimicking a power law dependence. As a 
result, it has been shown that ($a$) the electromagnetic coupling hits a 
Landau pole at as low a scale as $\Lambda \approx 40R^{-1}$, and ($b$) 
there is approximate (but not exact) unification of the three gauge 
coupling constants at an even lower scale $\Lambda \approx 20R^{-1}$. 
One therefore assumes that the low energy theory has a cutoff at either 
of these values, and phenomenological studies are made accordingly.  
This has been the standard practice in mUED studies over the past 
decade.

Of course, it is not only the gauge couplings that run faster in this 
model, but also the scalar self coupling $\lambda$. It has been 
shown~\cite{blitzkrieg} that if the self-coupling $\lambda = M_H^2/2v^2$ 
is less than 0.18 at the electroweak scale, then its renormalisation 
group evolution will inexorably drive it to zero at some high scale, at 
which point the electroweak vacuum will become unstable. Taking the 
experimental range 122~GeV~$\leq M_H \leq$127~GeV for the Higgs boson 
mass, we obtain $0.123 \leq \lambda \leq 0.133$, which is clearly below 
0.180. It follows that the electroweak vacuum in this model will indeed 
destabilise at some high scale, as, in fact, happens in the Standard 
Model itself at very high scales. The surprise lies in that fact 
that the `power law' running of $\lambda$ in the mUED model is so fast that 
the destabilisation takes place at a scale which is always below 
$6R^{-1}$. At this surprisingly low scale, new physics must come to the 
rescue, and hence the destabilisation scale can be treated as a cutoff 
for the mUED model.

\begin{figure}[h]
\centerline{\epsfig{file=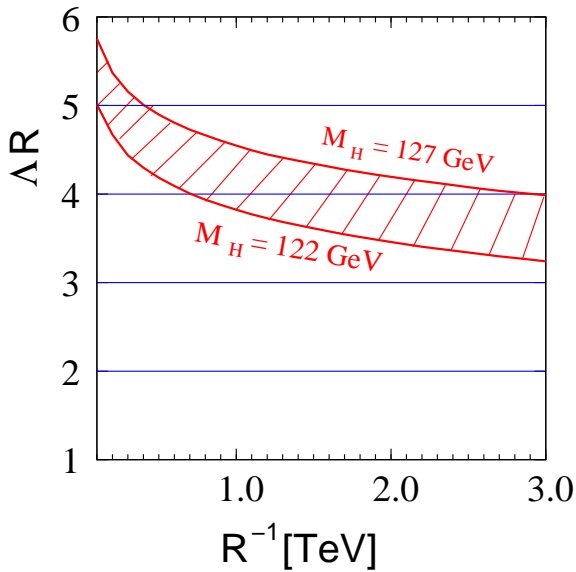,height=3.0in, width = 3.0in}}
\def\baselinestretch{1.1}
\caption{\footnotesize\sl Variation of $\Lambda/R^{-1}$, where $\Lambda$ 
is the cutoff induced by destabilisation of the electroweak vacuum, as a 
function of size parameter $R^{-1}$. The (red) hatched band represents 
variations in the Higgs boson mass from 122 -- 127 GeV, and horizontal 
(blue) lines represent KK levels.}
\def\baselinestretch{1.5}
\label{fig:cutoff}
\end{figure}

The exact value of the cutoff scale is determined by evaluating the 
running coupling constant $\lambda$ and determining where it 
vanishes~\cite{blitzkrieg}. The most important input parameters which 
determine this running are the mass of the Higgs boson ($M_H$) and the 
size parameter ($R^{-1}$), which is nothing but the inverse of the 
compactification radius of the extra dimension. The solid (red) lines in 
Figure~\ref{fig:cutoff} show the variation of the cutoff scale 
$\Lambda$, in units of $R^{-1}$, as a function of this size parameter 
$R^{-1}$, for two values of Higgs boson mass $M_H = 122, 127$~GeV (which 
represent the 3$\sigma$ experimental limits). The (red) hatching, 
therefore, represents all the intermediate values of $M_H$. Horizontal 
(blue) lines represent the different KK levels $n/R$, for $n = 
1,2,\dots,6$. Our results shown here correspond closely to similar results
shown in Ref.~\cite{blennow}.

Obviously, assuming tree-level masses, the number of KK modes with mass 
$M_n \approx n/R$ which can participate in any process will be given by 
the nearest integer lower than the solid (red) curve for a given value 
of $R^{-1}$. It is clear that this number can only vary between 3 and 5, 
and can never reach higher values such as 20 and 40 which used to be 
assumed earlier. Note that in generating Figure~\ref{fig:cutoff}, and 
subsequently, we have fixed the top quark mass at $m_t = 172.3$~GeV. 
Variation of the top quark mass between its experimentally allowed 
limits~\cite{top_mass} does result in some distortion of the curves, as 
the related Yukawa coupling plays a role in the running of the 
self-coupling $\lambda$. However, these distortions have very minor 
effects on the final conclusions of this article, and hence are not 
shown here.

In an earlier article~\cite{ADSR}, written at a stage when the new boson 
discovered at the CERN LHC had not yet been identified with any 
certainty as the Higgs boson, two of the present authors had shown that 
this low value of cutoff (i.e. small number of KK modes to sum over) 
leads to a compressed spectrum of KK modes of SM fields at any level $n 
\geq 1$, which presents serious difficulties for detection at the 
Tevatron and LHC. However, it was not possible to impose constraints on 
the model from the Higgs boson decay branching ratios, which were very 
imperfectly measured~\cite{Higgs_discovery} at that stage. Now, however, 
we have better experimental results on these branching 
ratios~\cite{ATLAS-EPF, CMS-EPF}, which, though not as precise or 
consistent between separate experiments as we would have liked them to 
be, have nevertheless reached a level where they are accurate enough to 
begin to constrain the mUED model~\cite{Belanger}. These constraints 
form the subject of the present study.

Before we go on to actually study the Higgs boson decay widths, however, 
it may be noted that bounds on the size parameter $R^{-1}$ quoted from 
hadron collider studies~\cite{collider_bounds} are generally based on 
expanded spectra arising when we sum KK levels up to $N = 20$ or even 
$N = 40$, which, as we have shown, is incompatible with stability of the 
electroweak vacuum. {\it We should set aside such hadron 
collider bounds on the mUED model}. The LEP bound $R^{-1} > 260$~GeV, 
obtained at $3\sigma$ from precision electroweak tests~\cite{ADSR}, may, 
however, be taken as a certainty. In a recent 
work~\cite{Edelhauser}, it has been shown that even if we sum up to 
5 KK levels, a lower bound of $R^{-1} > 720$~GeV at 95\% C.L. can be obtained 
by noting the non-observation by the CMS Collaboration of dilepton 
signals~\cite{CMSdilepton} arising from the decay of 
$n = 2$ resonances of the mUED model in the 7-8~TeV runs of the LHC. The 
purpose of the present study is, therefore, to ascertain if the existing data
on the Higgs boson decay channels can provide even better constraints.

Turning then, to the Higgs boson decays, the actual 
experimentally-measured quantities are the so-called {\it signal 
strengths}~\cite{ATLAS-EPF, CMS-EPF}. For a decay $H \to X\bar{X}$, the 
signal strength is defined by
\begin{eqnarray}
\mu_{X\bar{X}} & = & 
\frac{\sigma(pp \to H^0) \times {\cal B}(H^0 \to X\bar{X})}
{\sigma^{\rm (SM)}((pp \to H^0) \times {\cal B}^{\rm (SM)}(H^0 \to X\bar{X})}
\end{eqnarray}
where ${\cal B}(H^0 \to X\bar{X})$ is the branching ratio of the Higgs 
boson to an $X\bar{X}$ pair, and $\sigma(pp \to H^0)$ is the 
cross-section for single Higgs production at the LHC. The superscript 
(SM) denotes the SM prediction. Obviously, if the SM is the correct 
theory, then the experimental data will eventually converge on the 
results $\mu_{X\bar{X}} \simeq 1$ for all the channels $X$. On the other 
hand, deviations from unity will indicate new physics. As of now, the 
ATLAS and CMS Collaborations at CERN have measured signal strengths for 
$X\bar{X} = WW^\ast, ZZ^\ast, b\bar{b}, \tau^-\tau^+, \gamma\gamma$. Of 
these, the case $X\bar{X} = b\bar{b}$ is not very viable yet because of 
large errors. The other four have been measured with a better degree of 
precision. The results are given in Table~\ref{tab:mews} below.

\begin{table}[!htbp]
\begin{center}
\begin{tabular}{rcccc}
\hline 
      & $\mu_{WW}$             & $\mu_{ZZ}$  
      & $\mu_{\tau\tau}$       & $\mu_{\gamma\gamma}$      \\
\hline \\ [-3mm]
ATLAS & $0.99^{+0.31}_{-0.28}$ & $1.43^{+0.40}_{-0.35}$ 
      & $0.8 \pm 0.7$          & $1.55^{+0.33}_{-0.28}$    \\ [1mm]
CMS   & ~~$0.68 \pm 0.20$~~        & ~~$0.92 \pm 0.28$~~
      & ~~$1.10 \pm 0.41$~~        & ~~$0.77 \pm 0.27$~~           \\ [1mm]
\hline
\end{tabular}
\def\baselinestretch{1.1}
\caption{\footnotesize\sl ATLAS~\cite{ATLAS-EPF} and CMS~\cite{CMS-EPF} 
data on Higgs boson signal strengths, as reported in the summer of 2013. 
For $\mu_{\tau\tau}$ we use the March 2013 results of 
ATLAS~\cite{ATLAS-Moriond}.} 
\def\baselinestretch{1.5}
\label{tab:mews}
\end{center}
\vspace*{-0.3in} 
\end{table}

We now discuss how to predict the values of $\mu_{X\bar{X}}$ in the mUED 
model. Using the fact that the parton-level cross-section for gluon 
fusion $gg \to H^0$ is related to the decay width of $H^- \to gg$ by the 
linear relation
\begin{equation}
\sigma(gg \to H^0) = \frac{\pi^2}{8M_H^3} \Gamma(H^0 \to gg) \ ,
\end{equation}
we can rewrite the signal strength entirely in terms of decay widths as
\begin{eqnarray}
\mu_{X\bar{X}} & = & 
\frac{\Gamma(H^0 \to gg)}{\Gamma^{\rm (SM)}((H^0 \to gg)} \times 
\frac{\Gamma(H^0 \to X\bar{X})}{\Gamma^{\rm (SM)}(H^0 \to X\bar{X})} \times
\frac{\Gamma^{\rm (SM)}_H}{\Gamma_H} 
\label{eqn:singal} 
\end{eqnarray}
where
\begin{equation}
\Gamma_H = \sum_{X} \Gamma(H^0 \to X\bar{X})
\end{equation}
and all PDF-related effects (to leading order) in the cross-section may 
be expected to cancel in the ratio.  All we have to do, therefore, is to 
calculate the decay widths of the Higgs boson in the mUED model and the 
SM, and take the appropriate ratios. All the formulae relevant for these 
are available in the literature, but, for the sake of completeness and 
having a consistent notation, we list the most important formulae below.

In the SM, the decay width of the Higgs boson to a pair of leptons is 
given by~\cite{Barger}
\begin{equation}
\Gamma(H^0 \to \ell^+\ell^-) = 
\frac{\alpha(M_H)}{8\sin^2\theta_W} 
\frac{m_\ell^2}{M_W^2} M_H
\left(1 - \frac{4m_\ell^2}{M_H^2}\right)^{3/2} 
\end{equation}
where $\alpha(Q)$ is the running QED coupling at the mass scale $Q$.
The corresponding decay width to a pair of quarks is given by~\cite{Barger}
\begin{equation}
\Gamma(H^0 \to q\bar{q}) = 
\frac{3\alpha(M_H)}{8\sin^2\theta_W} 
\frac{m_q^2(M_H)}{M_W^2} M_H
\left(1 - \frac{4m_q^2}{M_H^2}\right)^{3/2}
\left\{1 + 5.67\frac{\alpha_s(M_H)}{\pi}\right\} 
\end{equation}
where the last factor represents the QCD corrections to the decay 
width~\cite{QCD_Hqq}, and the running quark mass is given 
by~\cite{running_mq}
\begin{equation}
m_q^2(M_H) = m_q^2 \left\{
\frac{\alpha_s(M_H)}{\alpha_s(m_q)} \right\}^{24/23}
\end{equation}
where $\alpha_s(Q)$ is the running QCD coupling at the mass scale $Q$.

The SM decay width of the Higgs boson to a $WW^\ast$ pair is given 
by~\cite{Marciano}
\begin{equation}
\Gamma(H^0 \to WW^\ast) = \frac{3\alpha^2(M_H)}{32\pi\sin^4\theta_W M_H} 
F(M_W)
\end{equation}
and that to a $ZZ^\ast$ pair by~\cite{Marciano}
\begin{equation}
\Gamma(H^0 \to ZZ^\ast) = \frac{\alpha^2(M_H)}{72\pi\sin^42\theta_W M_H}
\left( 63 - 120\sin^2\theta_W + 160\sin^4\theta_W \right)  F(M_Z)
\end{equation}
where
\begin{eqnarray}
F(M) = 
&-&\frac{1}{2}\left(1 - \frac{M^2}{M_H^2}\right)
\left(47M^2-13M_H^2+\frac{2M_H^4}{M^2}\right)
\nonumber \\
&-&3\left(M_H^2-6M^2+\frac{4M^4}{M_H^2}\right)\ln\frac{M^2}{M_H^2}
\nonumber \\
&+&3\left(M_H^2-8M^2+\frac{20M^4}{M_H^2}\right) 
\frac{M_H}{\sqrt{4M^2 - M_H^2}}
\cos^{-1} \frac{M_H}{2M}\left( 3 - \frac{M_H^2}{M^2} \right)
\end{eqnarray}
It is important to note that QCD corrections are significant only in the 
decay widths of the Higgs boson to quarks and can be neglected for all 
other decay modes. Likewise, the mUED contributions to the above decay 
modes is negligible, arising, as they do, from higher order effects 
which are severely suppressed by the heavy masses of the KK modes.

The decay modes which will be of most interest in the present work, are 
however, those that occur at the one-loop level in the SM, viz. the 
decays of the Higgs boson to a pair of gluons ($H^0 \to gg$) or a pair 
of photons ($H^0 \to \gamma\gamma$). Formulae for the partial decay 
widths in the SM are given in Ref.\cite{Barger}, and the extra 
contributions in the mUED, which occur at the same level in perturbation 
theory, are given in Ref.\cite{Petriello}. We list, below, these 
formulae in a common notation, with a couple of modifications to the 
formulae of Ref.\cite{Petriello}, which will be mentioned at the 
appropriate juncture.

The partial decay width of the Higgs boson to a pair of gluons is given by
\begin{eqnarray}
\Gamma(H^0 \to gg) & = &
\frac{\alpha(M_H)\alpha_s^2(M_H)}{72\pi^2\sin^2\theta_W}
\frac{1}{M_H^5 M_W^2} \mid\Omega^{\rm (SM)}_{gg} 
+ \Omega^{\rm (KK)}_{gg} \mid^2 \nonumber \\
& & \times
\left\{1 + 17.92\frac{\alpha_s(M_H)}{\pi} 
+ 156.8\frac{\alpha_s^2(M_H)}{\pi^2}
+ 467.7\frac{\alpha_s^3(M_H)}{\pi^3} \right\} 
\end{eqnarray} 
where the second line indicates the QCD corrections~\cite{QCD_Hqq} and 
the loop integral functions are given by
\begin{eqnarray}
\Omega^{\rm (SM)}_{gg} & = & 
\sum_q 3m_q^2\left\{ 2M_H^2 - (M_H^2 - 4m_q^2)f(m_q) \right\}  \\
\Omega^{\rm (KK)}_{gg} & = & \sum_q \sum_{n=1}^N 
3m_q^2\left\{ 4M_H^2 - (M_H^2 - 4m_{q,n,1}^2)f(m_{q,n,1}) 
- (M_H^2 - 4m_{q,n,2}^2)f(m_{q,n,2})\right\} \nonumber
\end{eqnarray}
where $m_{q,n,1}$ and $m_{q,n,2}$ are the two eigenvalues of the mass matrix 
\begin{equation}
{\cal M}_q^{(n)} = \left( \begin{array}{cc}  
m_{qL}^{(n)}  &   m_q  \\
m_q           &  -m_{qR}^{(n)} \end{array} \right)
\label{eqn:massmatrix}
\end{equation}
for the $n$'th level KK modes of the quarks, where
\begin{equation}
\left[m_{qL}^{(n)}\right]^2 = \frac{n^2}{R^2} + m_q^2 + \delta_{qL}^{(n)}
\qquad\qquad
\left[m_{qR}^{(n)}\right]^2 = \frac{n^2}{R^2} + m_q^2 + \delta_{qR}^{(n)}
\end{equation}
in terms of the radiative corrections $\delta_{qL}^{(n)}$ and 
$\delta_{qR}^{(n)}$~\cite{Matchev}. The function $f(m)$ is the usual 
loop integral~\cite{Barger}
\begin{equation}
f(m) = \left\{  \begin{array}{l}  
-2 \left( \sin^{-1} \frac{M_H}{2m} \right)^2 
\hskip 67pt  {\rm for}~~ m > \frac{M_H}{2} \\ [2mm]
-\frac{\pi^2}{2} \hskip 127pt {\rm for}~~ m = \frac{M_H}{2} \\ [2mm]
\frac{1}{2}\left( \ln\frac{M_H + \sqrt{M_H^2 - 4m^2}}{M_H - \sqrt{M_H^2 - 4m^2}}
- i\pi \right)^2 ~~{\rm for}~~ m < \frac{M_H}{2}  \end{array} \right.
\end{equation}

In using these formulae, we differ from Ref.~\cite{Petriello} in two ways: 
\vspace*{-0.2in}
\begin{enumerate}

\item we consider the sum over KK modes to terminate at $N$, which is 
the largest integer smaller than $\Lambda R$ as given in 
Fig.~\ref{fig:cutoff}, instead of summing to infinity, as was done in 
Ref.~\cite{Petriello}; and

\item we consider the splitting between mass eigenstates of KK modes of 
quarks at the level $n$, whereas Ref.~\cite{Petriello} assumed them to 
be degenerate. Of course, the fact that the off-diagonal terms in the 
mass matrix of Eqn.~\ref{eqn:massmatrix} are $m_q$ indicates that such 
splitting between these states as does occur will be perceptible only in 
the third generation.
  
\end{enumerate}
\vspace*{-0.2in}
In a similar vein, the partial decay width of the Higgs boson to a pair 
of photons is given by
\begin{equation}
\Gamma(H^0 \to \gamma\gamma) = \frac{\alpha^3(M_H)}{16\pi^2\sin^2\theta_W}
\frac{1}{M_H^5 M_W^2} \mid\Omega^{\rm (SM)}_{\gamma\gamma} 
+ \Omega^{\rm (KK)}_{\gamma\gamma} \mid^2
\end{equation} 
where the loop integral functions are given by
\begin{eqnarray}
\Omega^{\rm (SM)}_{\gamma\gamma} & = & \sum_q e_q^2 \omega^{\rm (SM)}_q
+ \sum_\ell e_\ell^2 \omega^{\rm (SM)}_\ell + \omega^{\rm (SM)}_W \nonumber \\
\Omega^{\rm (KK)}_{\gamma\gamma} & = & \sum_{n=1}^N \left[
\sum_q e_q^2 \omega^{(n)}_q + \sum_\ell e_\ell^2 \omega^{(n)}_\ell 
+ \omega^{(n)}_W \right]
\end{eqnarray}
in terms of~\cite{Barger} 
\begin{eqnarray}
\omega^{\rm (SM)}_q & = & 3m_q^2\left\{ 2M_H^2 - (M_H^2 - 4m_q^2)f(m_q) 
\right\}
\nonumber \\
\omega^{\rm (SM)}_\ell & = & 
m_\ell^2\left\{ 2M_H^2 - (M_H^2 - 4m_\ell^2)f(m_\ell) \right\} \nonumber \\
\omega^{\rm (SM)}_W & = & 
-3M_W^2\left\{ M_H^2 - (M_H^2 - 2M_W^2)f(M_W)\right\} - \frac{1}{2}M_H^4
\end{eqnarray}
and ~\cite{Petriello} 
\begin{eqnarray}
\omega^{(n)}_q & = & 3m_q^2\left\{ 4M_H^2 - (M_H^2 - 4m_{q,n,1}^2)f(m_{q,n,1}) 
- (M_H^2 - 4m_{q,n,2}^2)f(m_{q,n,2}^2)\right\} \nonumber \\
\omega^{(n)}_\ell & = & m_\ell^2
\left\{ 4M_H^2 - (M_H^2 - 4m_{\ell,n,1}^2)f(m_{\ell,n,1}) 
- (M_H^2 - 4m_{\ell,n,2}^2)f(m_{\ell,n,2}^2)\right\} \nonumber \\
\omega^{(n)}_W & = & -4M_W^2 M_H^2 
+ \left\{ 4M_W^2\left(M_H^2 - 2M_{W,n}^2\right)
- M_{W,n}^2 M_H^2 \right\}f(M_{W,n}) - \frac{1}{2}M_H^4
\end{eqnarray}
where the lepton mass eigenvalues $m_{\ell,n,1}$ and $m_{\ell,n,2}$ are, 
for all practical purposes, degenerate.

\begin{figure}[h]
\centerline{\epsfig{file=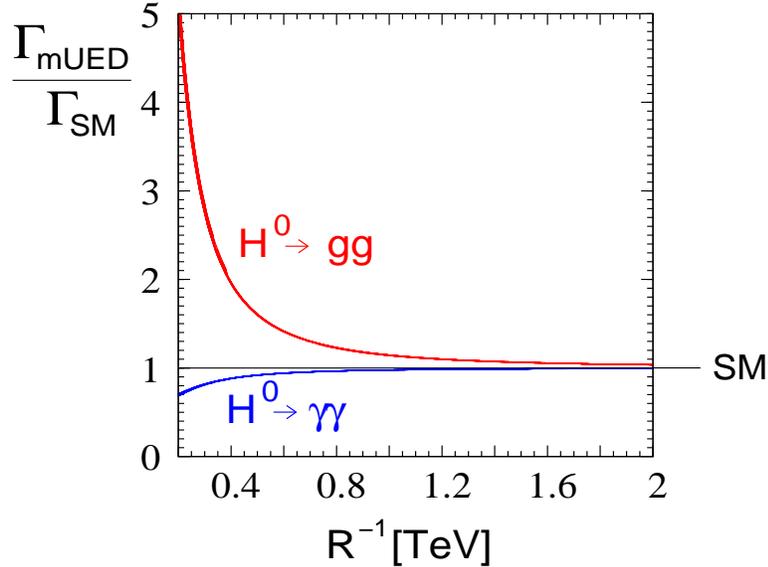,height=3.0in, width = 4.0in}}
\def\baselinestretch{1.1}
\caption{\footnotesize\sl Illustrating the effect of KK modes on the 
partial decay widths of $H^0 \to gg$ and $H^0 \to \gamma\gamma$. The 
former is always enhanced, while the latter is always suppressed, 
compared to the SM prediction.}
\def\baselinestretch{1.5}
\label{fig:ratios}
\end{figure}

Using these formulae, we can now find the signal strengths predicted in 
the mUED model as a function of the size parameter. To understand this 
behaviour, let us note the conclusion of Ref.~\cite{Petriello}, which 
remain qualitatively -- though not quantitatively -- true in our 
analysis as well. These may be summed up as follows.
\vspace*{-0.2in}
\begin{itemize}
\item The tree-level decay widths of the Higgs boson are practically the 
same in the SM and the mUED model.
\item The decay width of the Higgs boson to a pair of gluons is 
considerably enhanced in the mUED model, especially when $R$ is taken 
close to its lower experimental bound (see Figure~\ref{fig:ratios}).
\item The decay width of the Higgs boson to a pair of photons is 
suppressed in the mUED model, especially when $R$ is taken close to its 
lower experimental bound (see Figure~\ref{fig:ratios}).
\end{itemize}
\vspace*{-0.2in}

In our analysis, we obtain numerically different results from 
Ref.~\cite{Petriello} because of two reasons. In the first place, we 
note that the sum over KK modes in our case is truncated at values of 
$n$ between 3 and 5, whereas Ref.~\cite{Petriello} took the sum to 
infinity. As a result, we obtain significantly smaller mUED 
contributions. The second point is that because of this low cutoff, we 
are able to take $R^{-1}$ somewhat lower than what the earlier 
collider-based bounds permit us, and these lower values could then lead 
to larger mUED contributions.

If we take a closer look at Eqn.~(\ref{eqn:singal}), however, we see 
that there are more conflicting effects. The three channels with 
$X\bar{X} = WW^\ast, ZZ^\ast$ and $\tau\tau$ will all receive 
enhancements in the mUED model through the first factor on the right of 
Eqn.~(\ref{eqn:singal}). The second factor will be practically unity, as 
we have explained above. The third factor, however, will suppress the 
signal strength if there are large enough mUED contributions in the 
first factor. Owing to these opposed effects, the enhancement in signal 
strength is not as large as it might have been otherwise.

A curious fact worth noting is that the variation in the last factor 
arises only because we do not yet have an accurate measurement of the total 
decay width of the Higgs boson. If the Higgs boson decay width could be 
accurately determined from a line shape analysis, as was done for the 
$W$ and $Z$ bosons at LEP and Tevatron, then that result alone could 
have been used to constrain any new physics model. In the case of the 
$\gamma\gamma$ channel, the second factor on the right of 
Eqn.~(\ref{eqn:singal}) will be somewhat smaller than unity, as a result 
of which the signal strength will be somewhat more suppressed than in 
the other cases. It is therefore
difficult, in the mUED model, to predict large excesses in the partial 
width of $H^0 \to \gamma\gamma$. We reiterate, therefore, that the mUED 
enhancement in $H^0 \to gg$ and the suppression of $H^0 \to \gamma\gamma$ 
are both in agreement with the results of Ref.~\cite{Petriello}, though 
the actual deviations are much more modest in the present case --- a 
consequence of the small number of KK modes which contribute to these 
deviations.

\begin{figure}[ht]
\centerline{\epsfig{file=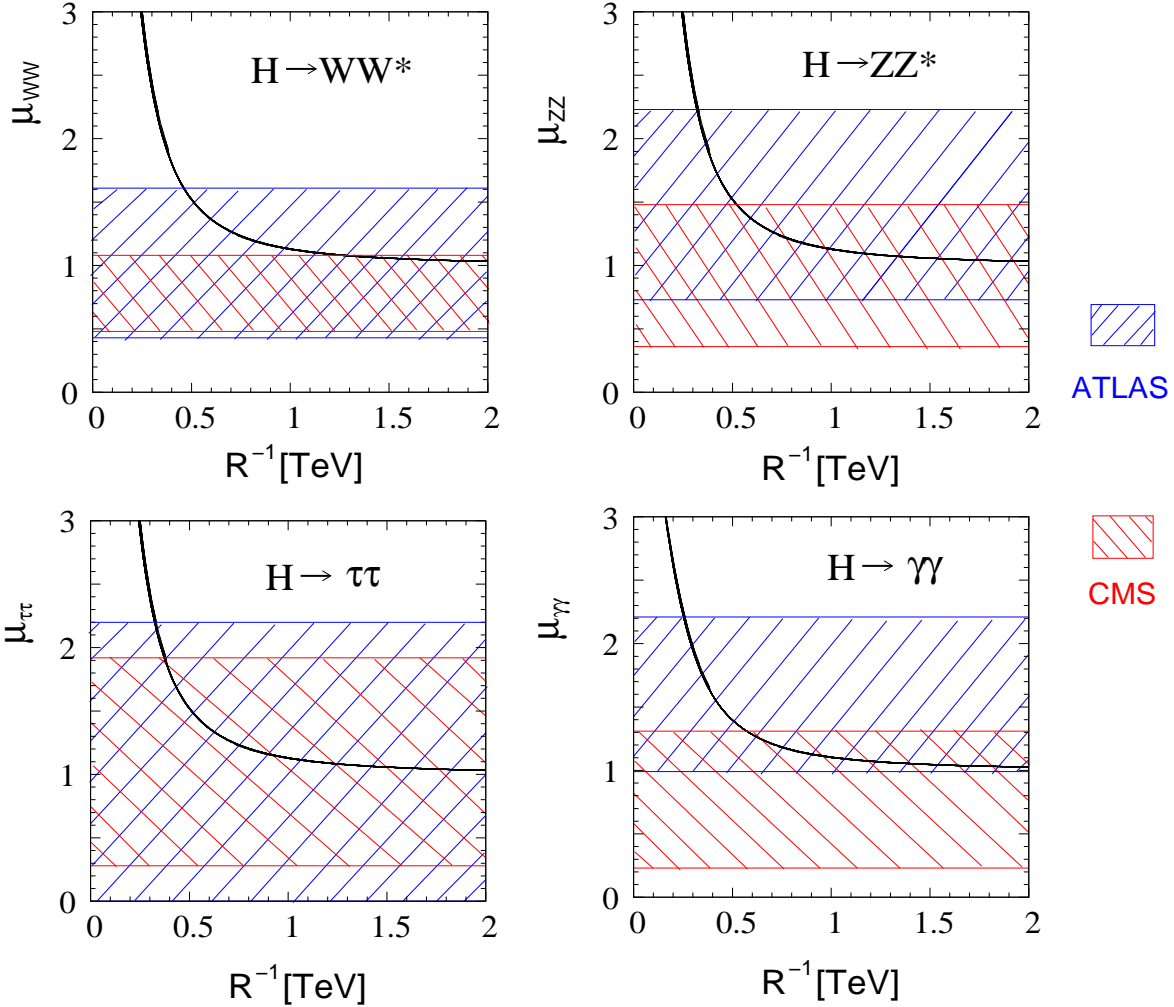,height=5.3in,width=6.2in}}
\vspace*{-0.15in}
\def\baselinestretch{1.1}
\caption{\footnotesize\sl Illustrating the variation with $R^{-1}$ of 
the signal strengths $\mu_{WW}$, $\mu_{ZZ}$, $\mu_{\tau\tau}$ and 
$\mu_{\gamma\gamma}$, as marked on the respective panels. The solid 
(black) lines show the mUED prediction, with their thickness 
representing the effect of varying the Higgs boson mass $M_H$ from $122 
- 127$~GeV. The oppositely-hatched regions (blue and red) denote, as 
indicated in the key on the right, the 95\% C.L. limits from the ATLAS 
and CMS Collaborations quoted in Table~\ref{tab:mews}.}
\def\baselinestretch{1.5}
\label{fig:signal}
\end{figure}
These diverse effects together contribute to the numerical results 
exhibited in Figure~\ref{fig:signal}. The four panels in this figure 
correspond to the four decays $H^0 \to WW^\ast, ZZ^\ast, \tau^+\tau^-$ 
and $\gamma\gamma$, as marked on each respective panel. The solid 
(black) lines represent the mUED predictions, and, as expected, these 
fall rapidly to the SM expectation $\mu_{X\bar{X}} = 1$ as $R^{-1}$ 
increases, in every case. The thickness of these lines indicates the 
effect of varying $M_H = 122 - 127$~GeV. It is clear from the figure 
that this is not a very significant effect\footnote{The effect of 
varying the top quark Yukawa coupling is sub-leading to this variation, 
which is why we do not show it at all in the present work.}. In fact, 
the solid (black) curves for $\mu_{WW}$, $\mu_{ZZ}$ and $\mu_{\tau\tau}$ 
are identical, since the only effect of introducing mUED lies in the 
first and last factors of Eqn.~\ref{eqn:singal}, which depend mainly on 
$\Gamma(H^0 \to gg)$. The solid (black) curve for $\mu_{\gamma\gamma}$ 
is clearly different, as one would expect. However, the reason for 
showing each signal strength separately lies in the fact that the 
experimental constraints are significantly different in each of these 
channels. For both the ATLAS and CMS data, the strongest constraints 
come, in fact, from the $WW^\ast$ channel. For a 125-126~GeV Higgs 
boson, these come out as $R^{-1} >$~463~GeV (1.3~TeV) for the ATLAS 
(CMS) results, which are far more restrictive than anything we can get 
from precision tests, and -- at least for the CMS data -- surpass the bounds
from dilepton channels~\cite{Edelhauser} by a factor close to 2.

\begin{figure}[h]
\centerline{\epsfig{file=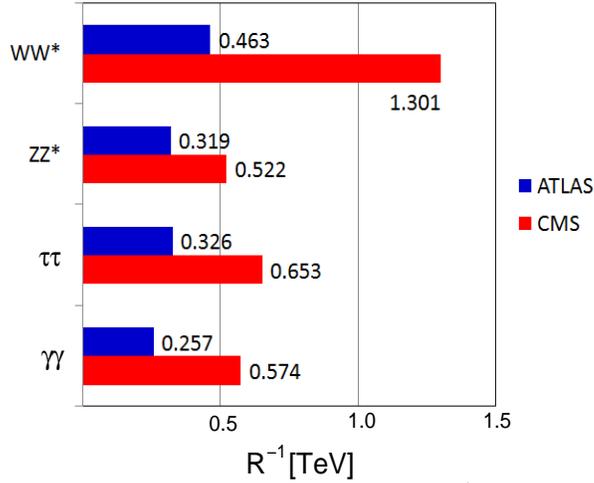,height=2.6in, width = 3.2in}}
\vspace*{-0.2in}
\def\baselinestretch{1.1}
\caption{\footnotesize\sl 95\% C.L. lower bounds (in TeV) on the size 
parameter $R^{-1}$ arising from four different Higgs boson decay 
channels. Numbers juxtaposed with the bars are the numerical value of 
the bounds.}
\def\baselinestretch{1.5}
\label{fig:bounds}
\end{figure}

95\% C.L. constraints from the other channels are illustrated, together 
with the $WW^\ast$ channel, in Figure~\ref{fig:bounds}, in the form of a 
bar graph.  It is apparent, even from Figure~\ref{fig:signal}, that the 
CMS data provide significantly stronger constraints, at this level, than 
the ATLAS data. In particular, if we consider the ATLAS data for $H^0 
\to \gamma\gamma$, where there appears to be an excess at the 1$\sigma$ 
level over the SM prediction, this appears to hint at lower values of 
$R^{-1}$, though -- as the graph shows -- large values of $R^{-1}$ are 
perfectly consistent with the 95\% C.L. limits. In view of the 
substantial differences between the two experimental results, it may be 
premature to read too much into these constraints, but it is clear that 
for the $WW^\ast$ channel, at least, we do find a reasonable level of 
consistency. Since this is the channel which provides the most stringent 
bounds on $R^{-1}$, these are perhaps the most acceptable among the four 
sets of constraints, at least at the present time.

In Figure~\ref{fig:bounds}, as mentioned above, we have shown a bar 
graph illustrating the individual 95\% C.L. constraints on $R^{-1}$ from 
each of these four channels. The upper (blue) and lower (red) bars 
represent bounds arising from the ATLAS and CMS data respectively. For 
the ATLAS data, the strongest constraint is from the $WW^\ast$ channel, 
but even the $ZZ^\ast$ and $\tau\tau$ channels are more restrictive than 
the LEP constraints. So far as the ATLAS data is concerned, obviously no 
useful constraint can be expected to arise from the $\gamma\gamma$ 
channel, but if the excess in this channel turns out to be a genuine 
feature, it will favour the mUED model (among other rival models) with a 
somewhat smaller value of $R^{-1}$. The CMS data, on the other hand, are 
much more restrictive.  While the
$WW^\ast$ channel pushes the lower bound to as high as 1.3~TeV, none of 
the other channels permit a value of $R^{-1}$ as low as 500~GeV, which 
is a substantial improvement over the LEP bound of 260~GeV, but is not
as restrictive as the dilepton bound obtained in Ref.~\cite{Edelhauser}.

The lower bound of $R^{-1} > 1.3$~TeV obtained from our computations 
represents a very strong constraint for the mUED model and would 
severely impact the direct searches planned for the 14~TeV run of the 
LHC. It is interesting, therefore, to ask how far these bounds can be 
relaxed if we consider the ATLAS and CMS data at the $3\sigma$ level 
rather than at 95\% confidence level. These bounds are presented in 
Table~\ref{tab:3sigma} below, and are naturally weaker, with the 
strongest bound lying at $R^{-1} > 685$~GeV, which is still a 
significant improvement over the precision tests\footnote{This is also 
definitely stronger than the $3\sigma$ bounds obtainable from dilepton 
signals, which would certainly lie around 600~GeV or below, if we go by 
the results quoted in Fig.~4 of Ref.~~\cite{Edelhauser}.}.

\begin{table}[!htbp]
\begin{center}
\begin{tabular}{rcccc}
\hline 
      & $\mu_{WW}$ & $\mu_{ZZ}$  & $\mu_{\tau\tau}$ & $\mu_{\gamma\gamma}$ \\
\hline \\ [-3mm]
ATLAS &   369      &   278       &   248       &  207   \\ [1mm]
CMS   &   685      &   413       &   306       &  402   \\ [1mm]
\hline
\end{tabular}
\def\baselinestretch{1.1}
\caption{\footnotesize\sl $3\sigma$ lower bounds (in GeV) on $R^{-1}$ 
using the ATLAS and CMS data from Table~\ref{tab:mews} and the signal 
strengths from Figure~\ref{fig:signal}.}
\def\baselinestretch{1.5}
\label{tab:3sigma}
\end{center}
\vspace*{-0.3in} 
\end{table}
If we further relax the constraints to the $5\sigma$ level, we find that 
the $WW^\ast$ channel data imply bounds on $R^{-1} > 280~(432)$~GeV from 
the ATLAS (CMS) data. Even with this very loose constraint, the lower 
bound of 432~GeV from the CMS data is still stronger than the LEP 
constraint. However, if we go by the conventional wisdom that $2\sigma$ 
deviations constitute a hint, $3\sigma$ deviations -- or the lack 
thereof -- constitute a bound, and $5\sigma$ is required for a 
discovery, then the stronger constraint $R^{-1} > 1.3$~TeV may be quite 
credible.

It is amusing to speculate on how these bounds might improve in the 
14~TeV run of the LHC --- under the somewhat pessimistic assumption that 
no deviations from the SM will be discovered. Estimates~\cite{Prospino} 
of the cross-section for $pp \to H^0$ at 8~TeV and 14~TeV indicate an 
enhancement in the cross-section by a factor around 2.5. Assuming that 
the integrated luminosity in the 14~TeV run will be as high as 1.5 
ab$^{-1}$, this represents an enhancement of 100 times over the 
statistics collected at 8~TeV. Thus, the number of Higgs boson events in 
the 14~TeV run will be around 250 times the number collected at the 
8~TeV run. If we concentrate on the $WW^\ast$ signal and assume that the 
errors will scale as the inverse square root of the number of Higgs 
boson decay events, then the error on the CMS measurement of $\mu_{WW}$ 
could go down as low as $0.012$. This is certainly an overestimate, 
since it does not take into account systematic effects, but it is 
probably safe to assume~\cite{Giacomelli} that the error could be as low as 5\%. Assuming, 
therefore, that we have a measured value $\mu_{WW} = 1.00 \pm 0.05$ 
(from either experiment, or from both combined), we immediately predict 
a 95\% C.L. limit $R^{-1} >1.58$~TeV, which would increase to 1.90~TeV 
if the integrated luminosity is doubled to 3~ab$^{-1}$. For such large 
values of $R^{-1}$, it is more or less sure that direct searches for 
mUED signals will fail, and even the LKP may become too heavy to explain 
the observed relic density of dark matter. In this admittedly 
pessimistic scenario, there will be no real motivation to study the mUED 
model any further.

Of course, we do not have any compelling reason to think that the above 
scenario is a true picture of the future. In fact, given the urgency 
with which an explanation of the composition of dark matter is required, 
we may well hope for just the reverse of this scenario, i.e. the 
observation of deviations in some of the Higgs boson partial decay 
widths in the 14~TeV run. In that case, we can reverse some of the 
arguments of the present study to show that a mUED explanation of such a 
deviation would be immediately available for some value of $R^{-1}$ in 
the range of $1 -2$~TeV.

To sum up, then, we have studied constraints on the mUED model from the 
measured Higgs boson signal strengths in the decays $H^0 \to WW^\ast$, 
$ZZ^\ast$, $\tau\tau$ and $\gamma\gamma$ channels. The mUED calculations 
have been carried out carefully, taking into account the fact that this 
model has a very low cutoff due to vacuum stability arguments. Even with 
the reduced effects due to this low cutoff, however, we find that the 
present CMS data can push the lower bound on the size parameter $R^{-1}$ 
of this model as high as 1.3~TeV at 95\% C.L. (or 685~GeV at $3\sigma$). 
ATLAS data are less restrictive, but in any case, do serve to push the 
value of $R^{-1}$ above about 500~GeV. All this represents an enormous 
improvement over the $3\sigma$ bound of around 260~GeV arising from 
precision electroweak tests at the LEP collider, as well as a factor close
to 2 greater than the 95\% dilepton bounds obtained from the early runs of 
the LHC. We then go on to argue that these signal strengths can be used to 
probe the mUED model up to $R^{-1} \approx 2$~TeV in the 14~TeV run of the 
LHC.

{\it Acknowledgements}: AD and SR would like to 
thank the Board of Research in Nuclear Studies, Government of India, for 
financial support under project no. 2013/37C/37/BRNS. AD also acknowledges 
partial financial support from the DRS programme at the Department of 
Physics, University of Calcutta. The work of AP is supported in part by 
the US Department of Energy grant no DE-FG-0204ER41036. He would also like to 
thank the DRS programme, Department of Physics, University of Calcutta, 
for hospitality during the summer of 2013, when part of this work was 
done.


\normalsize
\begin{thebibliography}{99}

\bibitem{Higgs_discovery} G.~Aad {\it et al.}  [ATLAS Collaboration], 
Phys.\ Lett.\ B {\bf 716}, 1 (2012) [arXiv:1207.7214 [hep-ex]]; 
S.~Chatrchyan {\it et al.}  [CMS Collaboration], Phys.\ Lett.\ B {\bf 
716}, 30 (2012) [arXiv:1207.7235 [hep-ex]]; TEVNPH Working Group (for 
the CDF, D0 Collaborations), Fermilab preprint FERMILAB-CONF-12-318-E, 
arXiv:1207.0449 [hep-ex] (2012).

\bibitem{Higgsless} S.~Dimopoulos and L.~Susskind, Nucl.\ Phys.\ B {\bf 
155}, 237 (1979); E.~Eichten and K.~D.~Lane, Phys.\ Lett.\ B {\bf 90}, 
125 (1980); C.~Csaki, C.~Grojean, H.~Murayama, L.~Pilo and J.~Terning, 
Phys.\ Rev.\ D {\bf 69}, 055006 (2004) [hep-ph/0305237].

\bibitem{strongly_coupled} B.~W.~Lee, C.~Quigg and H.~B.~Thacker, Phys.\ 
Rev.\ D {\bf 16}, 1519 (1977).

\bibitem{Landau_pole} W.~J.~Marciano, G.~Valencia and S.~Willenbrock, 
Phys.\ Rev.\ D {\bf 40} (1989) 1725; C.~F.~Kolda and H.~Murayama, JHEP 
{\bf 0007} (2000) 035 [hep-ph/0003170].

\bibitem{vacuum_stability} See M.~Sher, Phys.\ Rept.\ {\bf 179}, 273 
(1989), and references therein, for early work on the subject; for more 
recent work, see J.~Ellis, J.~R.~Espinosa, G.~F.~Giudice, A.~Hoecker and 
A.~Riotto, Phys.\ Lett.\ B {\bf 679}, 369 (2009) [arXiv:0906.0954 
[hep-ph]]; J.~Elias-Miro, J.~R.~Espinosa, G.~F.~Giudice, G.~Isidori, 
A.~Riotto and A.~Strumia, Phys.\ Lett.\ B {\bf 709}, 222 (2012) 
[arXiv:1112.3022 [hep-ph]]; G.~Degrassi, S.~Di Vita, J.~Elias-Miro, 
J.~R.~Espinosa, G.~F.~Giudice, G.~Isidori and A.~Strumia, JHEP {\bf 
1208}, 098 (2012) [arXiv:1205.6497 [hep-ph]]; F.~Bezrukov, 
M.Y.~Kalmykov, B.A.~Kniehl and M.~Shaposhnikov, JHEP {\bf 1210}, 140 
(2012); M.~Holthausen, K.S.~Lim and M.~Lindner, JHEP {\bf 1202}, 037 
(2012).

\bibitem{top_mess} S.~Alekhin, A.~Djouadi and S.~Moch, Phys.\ Lett.\ B 
{\bf 716} (2012) 214 [arXiv:1207.0980 [hep-ph]].
  
\bibitem{dark_matter} For a comprehensive discussion, see G.~Bertone, J. 
Silk, B. Moore, J. Diemand, J. Bullock, M. Kaplinghat, L. Strigari and 
Y. Mellier {\it et al.}, {\sl Particle Dark Matter: Observations, Models 
and Searches}, (Cambridge University Press, 2010).

\bibitem{dark_energy} P.~J.~E.~Peebles and B.~Ratra, Rev.\ Mod.\ Phys.\ 
{\bf 75}, 559 (2003) [astro-ph/0207347].

\bibitem{CP_violation} A.~Riotto and M.~Trodden, Ann.\ Rev.\ Nucl.\ 
Part.\ Sci.\ {\bf 49}, 35 (1999) [hep-ph/9901362].

\bibitem{SUSY} See, for example, S.~P.~Martin, {\sl A Supersymmetry 
Primer}, (in Kane, G.L. (ed.): ``Perspectives on supersymmetry II'', 
p.1), [hep-ph/9709356]; M.~Drees, R.~Godbole and P.~Roy, {\sl Theory and 
phenomenology of sparticles} (Hackensack, USA: World Scientific, 2004); 
H.~Baer and X.~Tata, {\sl Weak scale supersymmetry}, (CUP ,2006).

\bibitem{Appelquist} T.~Appelquist, H.C.~Cheng, B.A.~Dobrescu, Phys. 
Rev. {\bf D64}, 035002 (2001).

\bibitem{Hooper_Profumo} D.~Hooper and S.~Profumo, Phys.\ Rept.\ {\bf 
453}, 29 (2007) [hep-ph/0701197].

\bibitem{Matchev} H.-C.~Cheng, K.T.~Matchev, M.~Schmaltz, Phys. Rev.  
{\bf D66}, 036005 (2002); A.K.~Datta, K.C.~Kong, K.T.~Matchev, New J. 
Phys. {\bf 12}, 075017 (2010); B.~Bhattacherjee {\it et al}, Phys. Rev. 
{\bf D81}, 035021 (2010);

\bibitem{resonances} A.K.~Datta, K.C.~Kong, K.T.~Matchev, Phys. Rev. 
{\bf D72}, 096006 (2005); Erratum-ibid. {\bf D72}, 119901 (2005); 
B.~Bhattacherjee {\it et al}, Phys. Rev. {\bf D82}, 055006 (2010).

\bibitem{Gherghetta} K.R.~Dienes, E.~Dudas, T.~Gherghetta, Phys. Lett.  
{\bf B436}, 55 (1998) and Nucl. Phys. {\bf B537}, 47 (1999).

\bibitem{blitzkrieg} G.~Bhattacharyya {\it et al}, Nucl.Phys. {\bf 
B760}, 117 (2007).

\bibitem{blennow} M.~Blennow {\it et al}, Phys. Lett. {\bf B712}, 419 
(2012).

\bibitem{top_mass} [ATLAS Collaboration], ATLAS-CONF-2013-102 (2013).

\bibitem{ATLAS-EPF} [ATLAS Collaboration], CERN preprint 
CERN-PH-EP-2013-103 (2013).

\bibitem{CMS-EPF} [CMS Collaboration], CMS-PAS-HIG-13-005 (2013).

\bibitem{Belanger} G.~Belanger, A.~Belyaev, M.~Brown, M.~Kakizaki and 
A.~Pukhov, Phys.\ Rev.\ D {\bf 87}, 016008 (2013) [arXiv:1207.0798 
[hep-ph]]; U.~K.~Dey and T.~S.~Ray, Phys.\ Rev.\ D {\bf 88}, 056016 
(2013) [arXiv:1305.1016 [hep-ph]].

\bibitem{ADSR} A.~Datta and S.~Raychaudhuri, Phys.\ Rev.\ D {\bf 87},  
035018 (2013) [arXiv:1207.0476 [hep-ph]].

\bibitem{Edelhauser} L.~Edelhauser, T.~Fl\"acke and M.~Kramer, 
JHEP {\bf 1308}, 091 (2013) [arXiv:1302.6076 [hep-ph]].

\bibitem{CMSdilepton} CMS Collaboration], CMS-PAS-EXO-12-061 (2012).

\bibitem{collider_bounds} T.~Kakuda, K.~Nishiwaki, K.~-y.~Oda and 
R.~Watanabe, Phys.\ Rev.\ D {\bf 88}, 035007 (2013) [arXiv:1305.1686 
[hep-ph]].

\bibitem{ATLAS-Moriond} [ATLAS Collaboration], ATLAS-CONF-2013-034 
(2013).

\bibitem{Barger} V.~Barger and R.J.N.~Phillips, {\sl Collider Physics}, 
(Addison-Wesley, 2nd ed., 1997).

\bibitem{QCD_Hqq} M.~Schreck and M.~Steinhauser, Phys.\ Lett.\ B {\bf 
655}, 148 (2007) [arXiv:0708.0916 [hep-ph]].
  
\bibitem{running_mq} J. Beringer et al. (Particle Data Group), Phys. 
Rev. D86, 010001 (2012).

\bibitem{Marciano} W.~-Y.~Keung and W.~J.~Marciano, Phys.\ Rev.\ D {\bf 
30}, 248 (1984).

\bibitem{Petriello} F.~J.~Petriello, JHEP {\bf 0205}, 003 (2002) 
[hep-ph/0204067].

\bibitem{Prospino} W.~Beenakker, R.~Hopker and M.~Spira, hep-ph/9611232.

\bibitem{Giacomelli} See, for example, P.~Giacomelli, {\sl LHC: future 
measurements and reach}, Solvay Workshop on "Facing the Scalar Sector",
Brussels (May 2013).

  
\end{thebibliography}
\end{document}